\documentclass[aps,prl,superscriptaddress,twocolumn,eprint,nofootinbib]{revtex4-2}

\usepackage{amsmath,amsthm,amsfonts,amssymb,amscd}
\usepackage{mathtools}
\usepackage{fullpage}
\usepackage{lastpage}
\usepackage{hyperref}
\hypersetup{hidelinks}
\usepackage{physics}
\usepackage{graphicx}
\usepackage{subcaption}
\usepackage{bm}
\usepackage[dvipsnames]{xcolor}
\usepackage{float}
\usepackage[margin=0.55in]{geometry}
\usepackage[english]{babel}

\usepackage{comment}

\usepackage{tikz}

\usepackage{siunitx}

\usepackage[section]{placeins} 

\newcommand{\beq}{\begin{equation}}
	\newcommand{\eeq}{\end{equation}}

\begin{document}

	\title{Active nematic defects and epithelial morphogenesis}
	\author{Farzan Vafa}
	\affiliation{Department of Physics, University of California Santa Barbara, Santa Barbara, CA 93106, USA}
	\author{L. Mahadevan}
	\affiliation{School of Engineering and Applied Sciences, Harvard University, Cambridge, MA, USA}
	\affiliation{Departments of Physics, and Organismic and Evolutionary Biology, Harvard University, Cambridge, MA, USA}
	\date{\today}

	\begin{abstract}
		
		Inspired by recent experiments that highlight the role of nematic defects in the morphogenesis of epithelial tissues, we develop a minimal framework to study the dynamics of an active curved surface driven by its nematic texture. Allowing the surface to evolve via relaxational dynamics leads to a theory linking nematic defect dynamics, cellular division rates and Gaussian curvature. Regions of large positive (negative) curvature and positive (negative) growth are colocalized with the presence of positive (negative) defects. Applying this framework to the dynamics of cultured murine neural progenitor cells (NPCs) in an ex-vivo setting, we find that cells accumulate at positive defects and are depleted at negative defects. In contrast, applying this to the dynamics of a basal marine invertebrate \emph{Hydra} in an in-vivo setting, we show that activity stabilizes a bound $+1$ defect state by creating an incipient tentacle, while a bound $+1$ defect state surrounded by two $-1/2$ defects can create a stationary ring configuration of tentacles, consistent with observations.

	\end{abstract}
	
	\maketitle
	
\begin{figure}[t]
		\centering
		\begin{tikzpicture}
			\node[inner sep=0pt] (Q) at (0,0)
			{\includegraphics[width=.87\columnwidth]{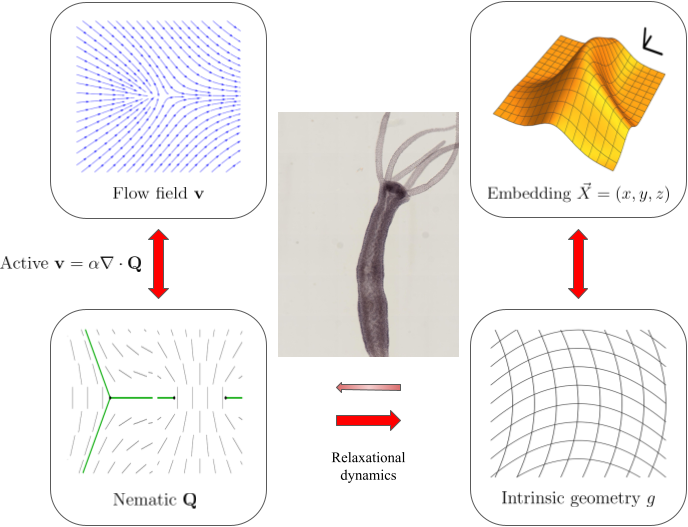}};
			\draw [black, line width=0.1em] (-.7,-3em) -- node[above,inner sep=0.1em, font=\tiny] {\SI{1}{mm}} (-.2,-3em);
		\end{tikzpicture}
		\caption{Schematic of our model. The activity through the flow field feeds into the model by stabilizing the nematic texture, and in particular the defects. The defects then induce growth and curvature for the intrinsic metric, which gets realized as the induced metric by the embedding. Image of \emph{Hydra} in the center, adapted from~\cite{drew} (approximate scale bar).}
		\label{fig:schematic}
	\end{figure}

	Morphogenesis, the origin of self-organized form in biology, results from the complex interplay of mechanical and biochemical processes~\cite{thompson1917on,barresi2020developmental}. To understand the dynamics of development, we need to complement our knowledge of the molecular constituents that unify many developmental programs across species with effective, coarse-grained theories that couple flows, forces, and self-regulation to generate shape and link them to testable experimental predictions~\cite{heisenberg2013forces,lecuit2017morphogenesis,gross2017how,collinet2021programmed}. At the cellular level, there are four geometric fields-- cell number, size, shape, and position--that vary in space and time that together are responsible for generating shape. In plant tissues, where cells do not change their relative positions, there has been much progress in linking molecular and cellular processes to tissue formation~\cite{meyerowitz1997genetic,coen2017on}. In animal epithelial tissues, the ability to tag and track thousands of cells in space and time~\cite{bosveld2012mechanical,tomer2012quantitative,krzic2012multiview} allow us to begin answering similar questions linking cellular processes to tissue shape ~\cite{farhadifar2007the,heer2017tension,irvine2017mechanical}. A particularly intriguing question in epithelial morphogenesis of nematically ordered cells is the role of topological defects in guiding or controlling morphogenesis, as has been uncovered in a few different systems, such as cell extrusion~\cite{saw2017topological,kawaguchi2017topological}, layer formation~\cite{copenhagen2020topological}, and the body shape of basal marine invertebrates such as {\em Hydra}~\cite{maroudas2020topological}. Here, we attempt to address the question of a minimal theory that focuses on the relaxational dynamics of the \emph{intrinsic} geometry of active epithelial surfaces, complementing work on evolving passive surfaces that has focused on/included allowing the \emph{induced} geometry to relax e.g.~\cite{lubensky1992orientational,seung1998defects,frank2008defects}, see~\cite{al-izzi2021active} for a recent review. In particular, we focus on the role of topological defects in determining how the intrinsic geometry of surfaces changes (Fig.~\ref{fig:schematic}), inspired by the observations in two different epithelial systems: cultured murine neural progenitor cells (NPCs)~\cite{kawaguchi2017topological} which show the onset of bulges/pits or sites of cellular apoptosis as a function of defect type, and \emph{Hydra}~\cite{maroudas2020topological} which shows a correlation between the sign of the local Gaussian curvature and the sign of the defect, as well as correlations between the formation of tentacles and the presence of bound arrays of defects.

	We model epithelial layers using the theory of active nematics, which consist of head-tail symmetric, elongated units that consume energy to move and do work on their surroundings, while still tending to align, locally generating nematic (apolar) order~\cite{marchetti2013hydrodynamics,doostmohammadi2018active,shankar2020topological}. Like their passive counterparts, active nematics exhibit singular distortions, i.e. {topological} defects which interrupt the nematic order (see \emph{Topological defects} of the SI for a brief review of nematic topological defects).
	
	A minimal model that couples the relevant degrees of freedom in an active system must allow for spatio-temporal variations in the two-dimensional nematic tensor $Q^{\mu\nu}$, a velocity field $v^a$ determined by an activity parameter which we will denote by $\alpha$ and which is controlled by the local nematic field, and the intrinsic 2D metric $g_{\mu\nu}$ (which can be deduced from tissue geometry), in addition to the extrinsic geometry of the sheet. For simplicity, here we will assume that the extrinsic geometry is harnessed to the intrinsic geometry, and further assume that the dynamics of the other fields are controlled by the gradient descent associated with variations in the free energy along with active flow dynamics. There are two main contributions to the free energy that we consider: (i) a contribution due to the nematic tensor $Q^{\mu\nu}=A[\hat n^\mu \hat n^\nu - \frac{1}{2} \delta^{\mu\nu}]$, where $A$ is the magnitude of the nematic order, and $\hat n^\mu$ is the local director field (ii) a contribution purely due to the metric $g_{\mu\nu}$, so that the total free energy $\mathcal F$ is the sum of contributions from the nematic field as well as from the intrinsic metric, with $\mathcal F = \mathcal F_Q + \mathcal F_g$.  
	
	For the nematic order parameter, the two-dimensional Landau-de Gennes free energy~\cite{gennes1993the}, $\mathcal F_Q$, in its covariant form can be written as
	\begin{align}
		\mathcal F_Q &= \int d^2x\sqrt{g}[Kg_{\beta\delta}\nabla_\alpha Q^{\alpha\beta} \nabla_{\gamma}Q^{\gamma \delta} \nonumber\\ 
		&\qquad + \frac{1}{4}\epsilon^{-2}(1 - 4g_{\beta\gamma} g_{\alpha\delta} Q^{\alpha\beta} Q^{\gamma\delta})^2] \nonumber\\
		&= \int d^2x \sqrt{g}[K\Tr[(\nabla\cdot Q)^2] +\frac{1}{4}\epsilon^{-2}(1 - 4\Tr [Q^2])^2]
		\label{eq:F_p}
	\end{align}
	where $g_{\mu\nu}$ is the metric and $\nabla_\alpha$ is the covariant derivative associated with it. Here $K (>0)$ is the elasticity in the single-elastic Frank constant approximation, and the last term governs the isotropic-nematic transition, with $\epsilon$ controlling the microscopic nematic correlation length.\footnote{One can also add $\beta R \Tr[Q^2]$ to $\mathcal F_Q$, where $R$ is the scalar curvature. This term will be generated depending on the choice of the two Frank constants in the non-flat geometry, and for simplicity, we will ignore it for now.}
	
	Inspired by the gradient dynamics of Ricci flow~\cite{hamilton1982three}, we can write the free energy due to variations in the metric $\mathcal F_g$ as	
	\beq \mathcal F_g = \int d^2x \sqrt{g}[K_\varphi R \varphi -\lambda], \eeq
	where $\sqrt{g} = \exp(\varphi)$, with $\varphi$ being a scalar field, $K_\varphi(>0)$ is an elastic constant penalizing changes in the Gauss curvature $R$, and $\lambda = \lambda(t)$ is the term accounting for growth which we can use to enforce the condition that the surface area does not change. 
	
	Then the dynamics of the nematic and metric fields associated with gradient descent and advection by a non-equilibrium flow $v^c$ yields
	\begin{align}
		\partial_t {Q}^{\alpha\beta} &= -v^c  \nabla_c {Q}^{\alpha\beta}+ [Q,\Omega ]^{\alpha\beta} -\gamma_Q^{-1}g^{\alpha\mu}g^{\beta\nu} \frac{1}{\sqrt{g}}\frac{\delta \mathcal F}{\delta Q^{\mu\nu}} \label{eq:dQdt}\\
		\partial_t g_{\alpha\beta} &= -(\nabla_\alpha v^c)g_{c\beta} -(\nabla_\beta v^c)g_{c\alpha}  - \gamma_\varphi^{-1} \frac{1}{\sqrt{g}} \frac{\delta \mathcal F}{\delta g^{\alpha \beta}} \label{eq:dgdt}\;,
	\end{align}
	with $\Omega_{ab} = (\nabla_a v_b - \nabla_b v_a)/2$ the vorticity, and $\gamma_Q$ and $\gamma_\varphi$ are the viscous coefficients for the dynamics of $Q^{\alpha\beta}$ and $g_{\alpha\beta}$, respectively, with units of radians$^2$/time.
	
	Closure of the system  \eqref{eq:dQdt}-\eqref{eq:dgdt} requires an equation for the active velocity field generated by  the active stress $\sigma^{\alpha\beta}$, which we assume is proportional to $Q^{\alpha\beta}$, that is, $\sigma^{\alpha\beta} = \tilde \alpha Q^{\alpha\beta}$ \cite{simha2002hydrodynamic,marchetti2013hydrodynamics}. Here  $\tilde\alpha$ has units of energy density, with $\tilde \alpha > 0$ ($\tilde \alpha < 0$) corresponding to contractile (extensile) activity. In the overdamped,  biologically relevant limit, assuming that flow arises due to the balance of active forces with the substrate friction $\mu$, we write
	\beq v^c = \alpha \nabla_\nu Q^{\nu c}\label{eq:v}\eeq
	with $\alpha = \tilde\alpha/\mu$ being what we refer to as the activity. Equations \eqref{eq:dQdt}-\eqref{eq:v} are a closed set of nonlinear partial differential equations that dictate the evolution of the nematic field $Q^{\alpha\beta}$ and the intrinsic geometry $g_{\alpha\beta}$ as a function of the activity $\alpha$, when complemented by appropriate initial and boundary conditions. We note that our theory is related to but simpler than recent phase field models for active deformable shells \cite{mietke2019self,metselaar2019topology} that account for both the induced and the intrinsic geometry of the manifolds.

	To make progress in a minimal setting for epithelial morphogenesis, we choose  2D isothermal (conformal) complex coordinates $z$ and $\bar z$ such that~\cite{david2004geometry} 
	\beq ds^2 = g_{z\bar z} dz d\bar z + g_{\bar z z}  d\bar z dz = 2g_{z\bar z} |dz|^2 = e^{\varphi} |dz|^2\eeq
	and assume that the metric remains diagonal in these coordinates for all time. Furthermore, since the nematic tensor $Q^{\mu\nu}$ is a traceless real bivector, we can write its components $Q = Q^{zz}$, $\bar Q = Q^{\bar z \bar z}$, and $Q^{z\bar z}$, with $Q^{z\bar z} = 0$, and  $Q = (\bar Q)^*$.	In these coordinates, $\mathcal F_Q$ and $\mathcal F_g$ take the form
	\begin{align}
		\mathcal F_Q &= \int d^2z\sqrt{g}[2Kg_{z \bar z}\nabla_z Q^{zz} \nabla_{\bar z}Q^{\bar z \bar z} + 2K'g_{z \bar z}\nabla_{\bar z} Q^{zz} \nabla_z Q^{\bar z \bar z} \nonumber\\
		&\quad + \frac{1}{4}\epsilon^{-2}(1 - 4g_{z \bar z}g_{z \bar z}Q^{zz}Q^{\bar z \bar z})^2] \nonumber\\
		&= \int d^2z \sqrt{g}[2K|\nabla_z Q|^2 + 2K'|\nabla_{\bar z} Q|^2 + \frac{1}{4}\epsilon^{-2}(1 - 4|Q|^2)^2] \\
		\mathcal F_g &= \int d^2z \sqrt{g}[K_\varphi R\varphi - \lambda]
	\end{align}
	where $Q = Q^{zz}$ and $\bar Q = Q^{\bar z \bar z}$, $|\cdot|$ is defined in terms of the metric, the covariant derivatives used are $\nabla_z Q^{zz} = \partial Q + 2(\partial\varphi)Q$ and $\nabla_{\bar z} Q^{zz} = \bar\partial Q$, and the scalar curvature $R = -4e^{-\varphi}\partial\bar\partial\varphi$. We note that the asymmetry in the appearance of $\partial\varphi$ between $\nabla_z Q$ and $\nabla_{\bar z}$ is the underlying reason behind asymmetry in cell growth near defects: cells accumulate at positive defects and deplete at negative defects.\footnote{Here we have added a term $\propto |\nabla_{\bar z} Q|^2$ to $\mathcal F_Q$, which is equivalent to adding a term $\propto R |Q|^2$ to $\mathcal F_Q$.}
	
	In terms of the eight independent parameters: $K$, $K'$, $\epsilon$, $\alpha$, $K_\varphi$, $\gamma_Q$, $\gamma_\varphi$, and the system size $L$, we can define a number of natural length and time scales:
	the nematic coherence length $\xi = \sqrt{K+K'}\epsilon$, which is proportional to the defect core radius, the geometric coherence length $\ell_\varphi = \sqrt{K_\varphi}\epsilon$, and
	the nematic response to curvature length $\ell_{R,Q} = \sqrt{|K - K'|}\epsilon$; and the relaxation times of $Q$ and $\varphi$, $\tau_Q = \gamma_Q \epsilon^2$ and $\tau_\varphi = \gamma_\varphi L^2/K_\varphi$.
	This leads to the following dimensionless quantities: $\xi/\ell_\varphi$, the ratio of coherence lengths for the nematic field and intrinsic geometry ($<1$ because extrinsic geometry variations occur on scales large compared to the nematic defect core size); $\tau_\varphi/\tau_Q = (\gamma_\varphi/\gamma_Q)(L/\ell_\varphi)^2$ 
	($\gg 1$ because we assume that the long wavelength extrinsic geometry relaxes slowly compared to the local nematic order);
	$K/K'$, the ratio of the two different types of nematic elastic deformations ($\sim 1$); and $K/(\gamma_Q\alpha)$, the ratio of passive to active stresses, which can be large or small.

	\begin{figure}[t]
		\subcaptionbox{}
		{\includegraphics[width=.4\textwidth]{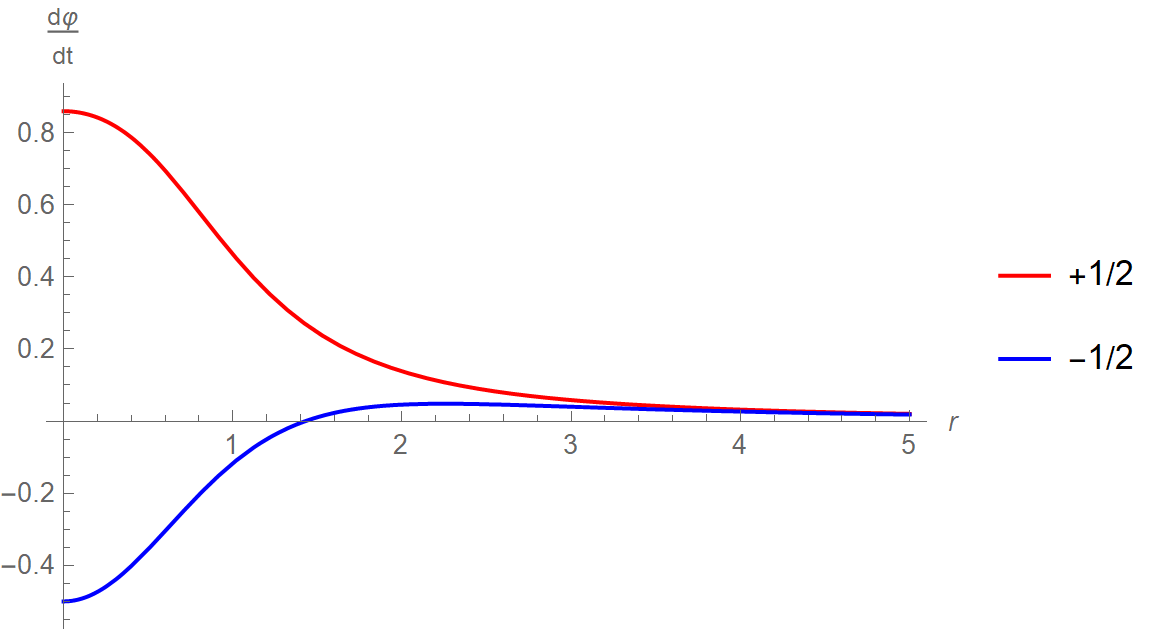}}

		\subcaptionbox{}
		{\includegraphics[width = .4\textwidth]{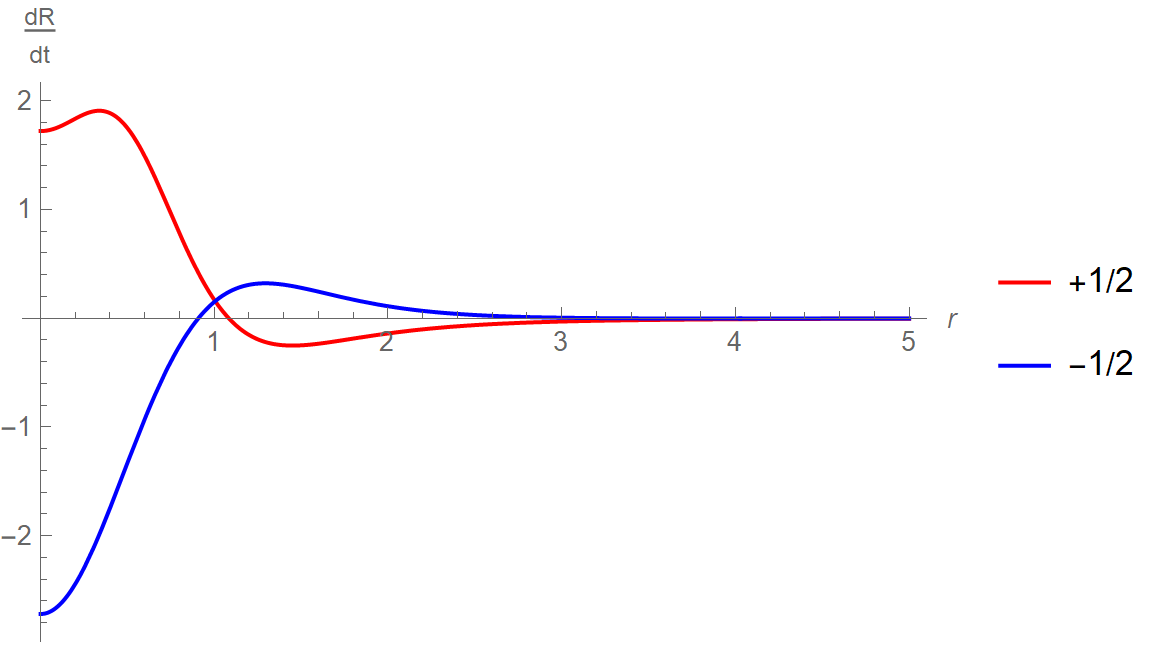}}
			\begin{picture}(0,0)
			\put(-120,190){\includegraphics[width=.17\textwidth]{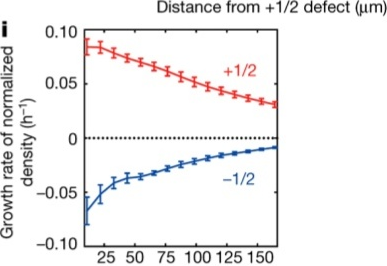}}
		\end{picture}
			\begin{picture}(0,0)
			\put(-120,100){\includegraphics[width=.17\textwidth]{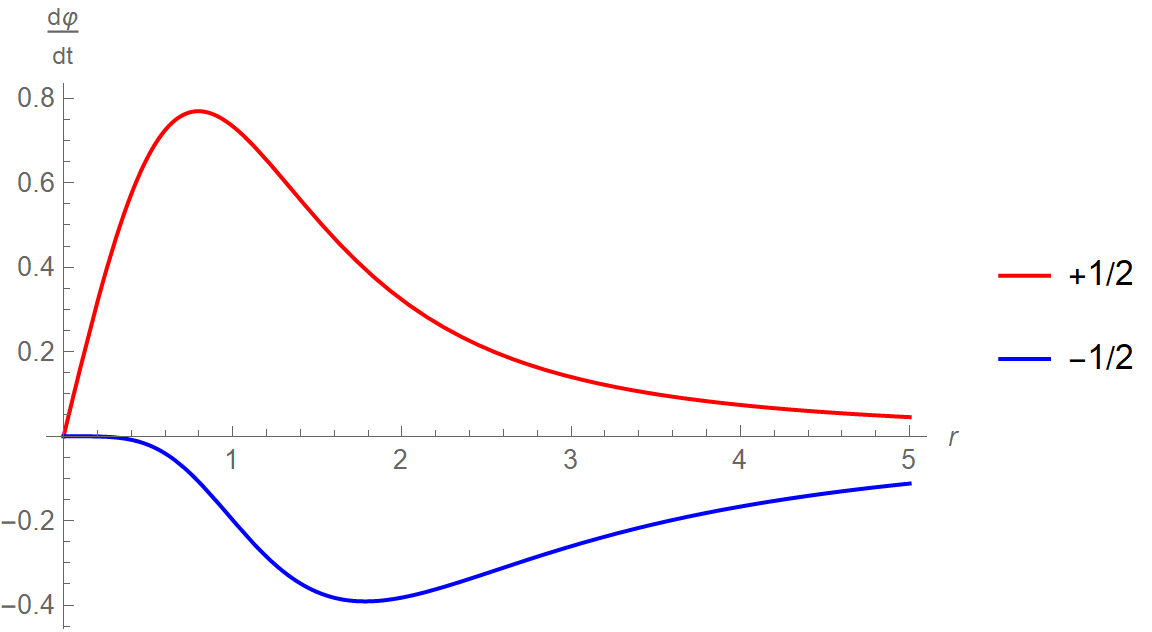}}
		\end{picture}
			\begin{picture}(0,0)
			\put(-120,0){\includegraphics[width=.17\textwidth]{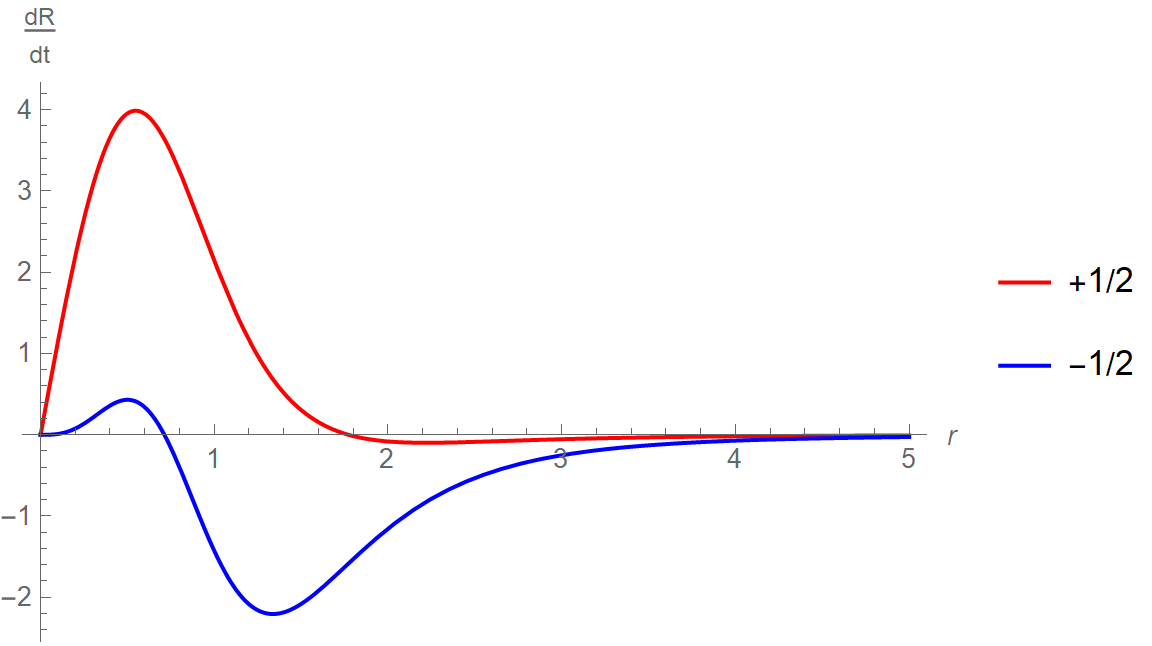}}
		\end{picture}
		\caption{Plots of (a) $\frac{d\varphi}{dt}$ and (b) $\frac{dR}{dt}$   for a single $+1/2$ (in red) and a single $-1/2$ defect (in blue) following from Eqs.~\eqref{eq:dphi+dt}-\eqref{eq:dphi-dt}, where $R = -4e^{-\varphi}\partial\bar\partial\varphi$. $\varphi$ and $R$ grow (shrink) for a $+1/2$ ($-1/2$) defect. Top inset: for comparison of $\frac{d\varphi}{dt}$, corresponding figure (growth rate of normalized cell density) adapted from~\cite{kawaguchi2017topological}. {Middle and bottom insets: corresponding plots of active contribution for $\frac{d\varphi}{dt}$ and  $\frac{dR}{dt}$ for $\phi=0$.
		In all of the plots, parameters used are	$K = 1$, $K' = 0$, $\epsilon=1$, and $\alpha=1$.}
		}
		\label{fig:initialDynamics}
	\end{figure}

	Before moving towards understanding the role of active defects in driving the intrinsic geometry of the surface, we first consider a passive nematic, when $\alpha = 0$, so that the dynamics for $Q$ and $\varphi$ in isothermal conformal coordinates can be written as 
	\begin{align}
		\gamma_Q\partial_t Q  &= 2Kg^{z\bar z}\nabla_{\bar z}\nabla_z Q + 2K'g^{z\bar z}\nabla_z\nabla_{\bar z} Q \nonumber\\
		&\quad + 2\epsilon^{-2} (1 - 4|Q|^2)Q \label{eq:dQComplexdt}\\
		\gamma_\varphi\partial_t\varphi & = -K_\varphi R + 4K|\nabla_z Q|^2 + 4Kg_{z\bar z}(Q\nabla_z\nabla_{\bar z} \bar Q + \bar Q \nabla_{\bar z}\nabla_z Q) \nonumber \\
		&\quad - 4K'|\nabla_{\bar z} Q|^2 - \frac{1}{2}\epsilon^{-2} (1 - 4|Q|^2)(1 - 20|Q|^2) {+ \lambda}, \label{eq:dphidt}
	\end{align} 
	where the new covariant derivative terms are $\nabla_{\bar z} \nabla_z Q = \bar\partial\partial Q + 2(\bar\partial\partial\varphi)Q + 2\partial\varphi\bar\partial Q$ and $\nabla_z \nabla_{\bar z} Q= \partial\bar\partial Q + 2\partial\varphi\bar\partial Q$.
	
	In the neighborhood of $\pm 1/2$ defects, we denote $\varphi^{\pm}$ and $Q^\pm$ as the local geometry and nematic field, respectively. If we start with a flat configuration, with $\varphi = 0$, Eq.~\eqref{eq:dphidt} describing the evolution of
	$\varphi$ in the neighborhood of a defect  simplifies to
	\begin{align}
		\gamma_\varphi\partial_t\varphi^+ &=  4K|\partial Q^+|^2 + 2K(Q^+\partial\bar\partial \bar Q^+ + \bar Q^+ \bar\partial\partial Q^+) \nonumber \\
		& \quad -4K'|\bar \partial Q^+|^2 - \frac{1}{2}\epsilon^{-2}(1 - 4|Q^+|^2)(1 - 20 |Q^+|^2) + \lambda \label{eq:dphi+dt}\\
		\gamma_\varphi\partial_t\varphi^- &=  4K|\partial Q^-|^2  + 2K(Q^-\partial\bar\partial \bar Q^- + \bar Q^- \bar\partial\partial Q^-) \nonumber \\
		&\quad -4K'|\bar \partial Q^-|^2 - \frac{1}{2}\epsilon^{-2}(1 - 4|Q^-|^2)(1 - 20 |Q^-|^2) + \lambda\label{eq:dphi-dt}
	\end{align}
	Now noting that $Q^+ = (Q^-)^*$ and that in the vicinity of the positive (negative) defect core  $\bar\partial Q^+$ ($\partial Q^-$) $= 0$ leads to
	\beq \gamma_\varphi \partial_t \varphi^+ - \gamma_\varphi \partial_t \varphi^- = 4K|\partial Q^+|^2 + 4K'|\bar\partial Q^-|^2 >0. \eeq
	In the absence of net surface growth, this implies that $\varphi$ will increase at a $+1/2$ defect and decrease near a $-1/2$ defect. Interpreting $\varphi$ as the log of the cell density (since the Gauss curvature $R = -4e^{-\varphi}\partial\bar\partial\varphi$), we expect the cell density to increase (decrease) at plus (minus) defects, which means we can interpret as cells accumulating (depleting) at the defects.

	\begin{figure}[t]
		\centering
		\begin{tikzpicture}
			\node[inner sep=0pt, label=below left:{(b)}] (phi) at (0,-7)
			{\includegraphics[width=\linewidth]{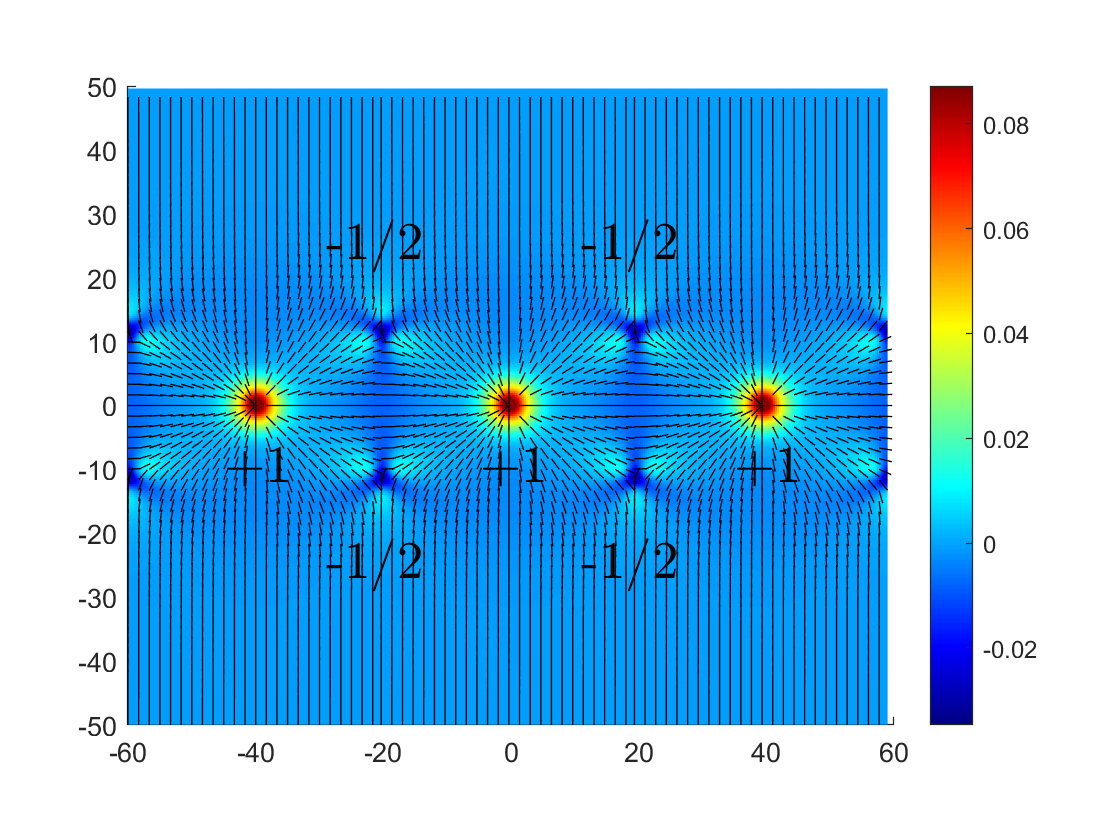}};
			\node[inner sep=0pt, label=below left:{(a)}] (Q) at (0,0)
			{\includegraphics[width=\linewidth]{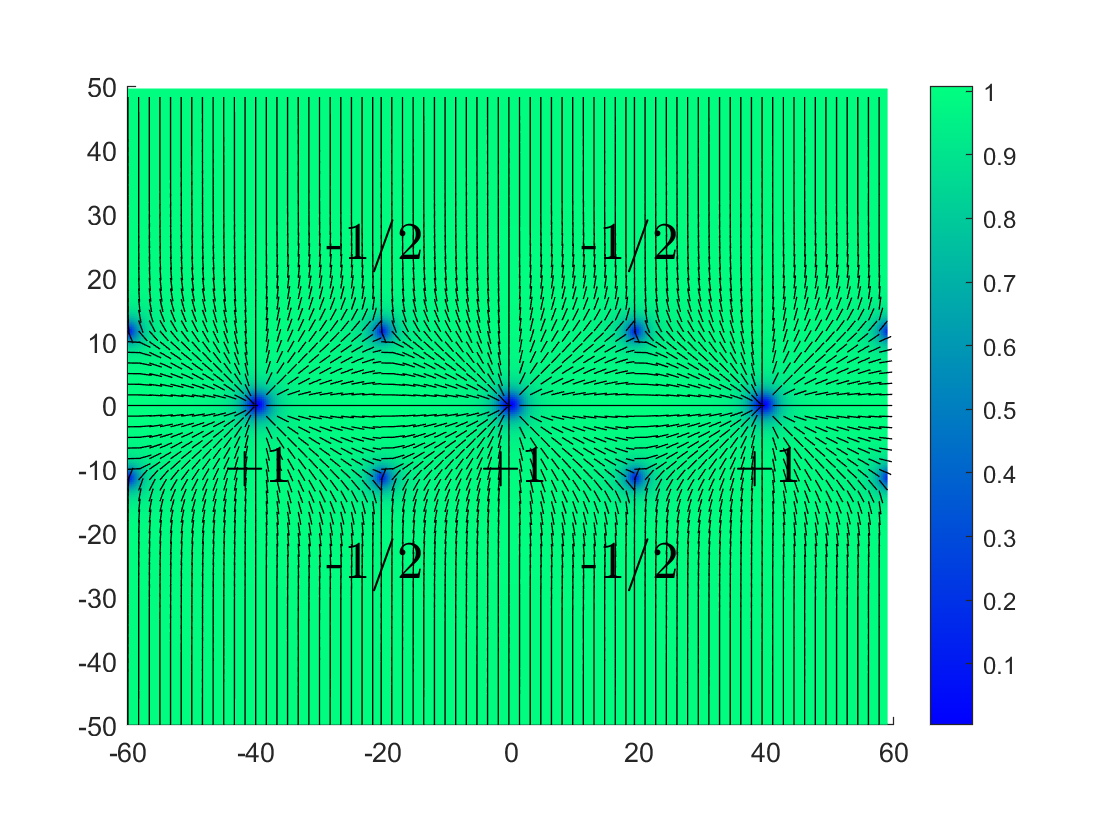}};
			\node[inner sep=0pt] (profile) at (-.45,-3.5)
			{\includegraphics[width=.4\linewidth]{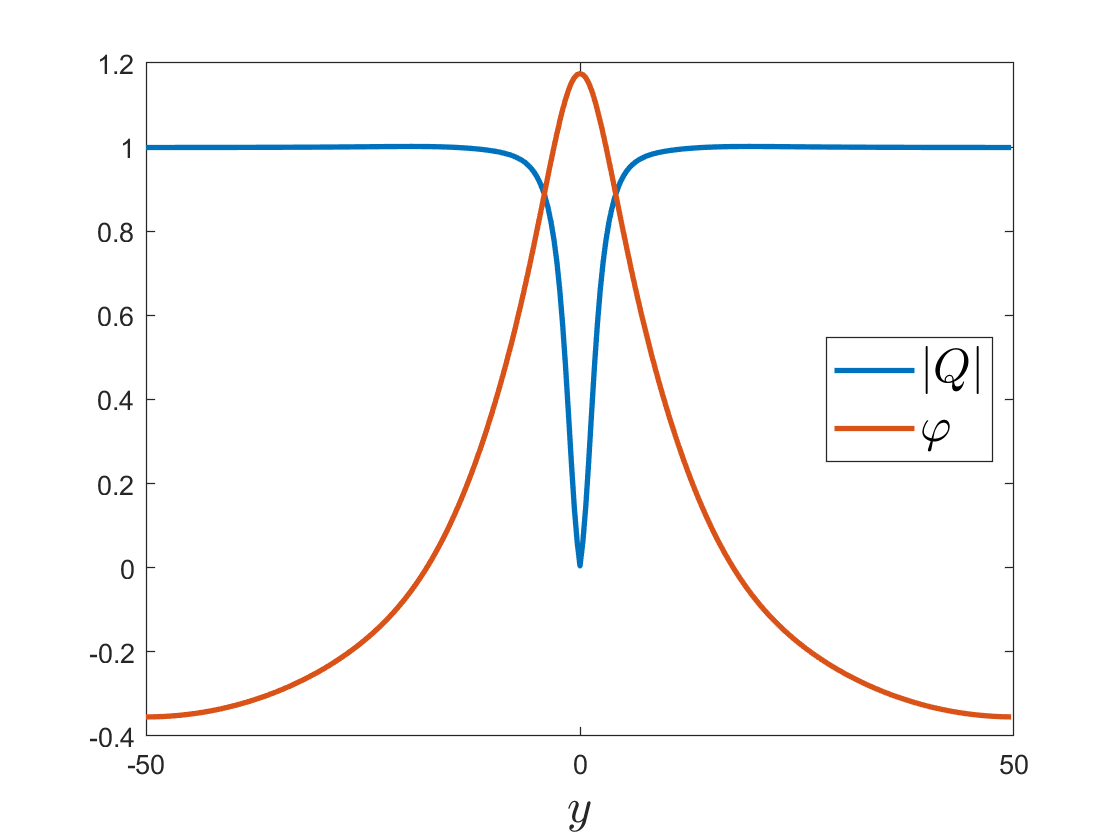}};
			\draw[thick] (-.9,-.7) rectangle (.1,.5);
			\draw[thick] (-.9,-7.6) rectangle (.1,-6.4);
			\draw[->,thick] (-.4,-1) -- (-.4,-1.9);
			\draw[<-,thick] (-.4,-5.2) -- (-.4,-6.1);
		\end{tikzpicture}
		\caption{We numerically integrate equations~\eqref{eq:dQComplexdt}-\eqref{eq:dphidt} (with the substitution $\partial_t \to D_t$) to obtain
			steady state plots of (a) the magnitude of the nematic order parameter $|Q|$ and (b) the curvature density (given by $-4\partial\bar\partial \varphi$). We note that the sign of the curvature correlates with the sign of the defect, and that the defect configuration is a lattice of $+1$ bound states separated by pairs of $-1/2$ defects. 
			In the inset, we show the profile of the nematic order $|Q|$ (blue) and $\varphi$ (red) along the $y$-axis. The profile of $|Q|$, which is dictated by the nematic coherence length, is smaller than the width of the profile of $\varphi$ since $\ell_\varphi > \xi$. Parameters for simulations: $\alpha=-0.8$, $K=1$, $K' = 0$, $\gamma_Q = \gamma_\varphi = 1$, $K_\varphi=4$, and $\epsilon=2$, in terms of which $\xi = 1$, $\ell_{R,Q} = 1$, and $\ell_\varphi = 2$. See text and \emph{Numerical methods} of SI for details.
		}
		\label{fig:latticePlots}
	\end{figure}
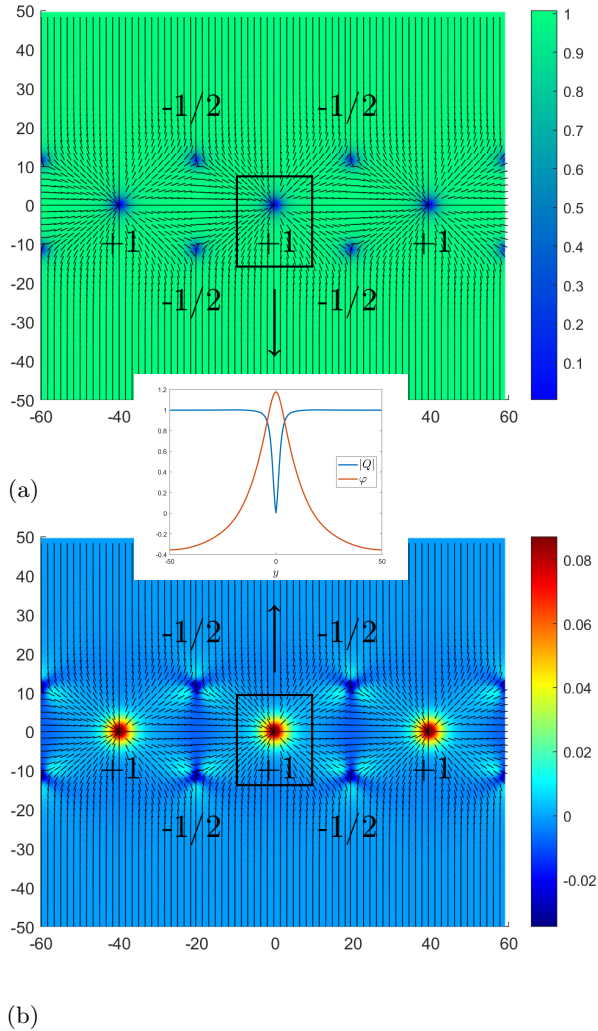
	
	In Fig.~\ref{fig:initialDynamics}, we compare this result with observations from two different experiments.   In the top panel is the initial profile of $\varphi$ at $t=0$ from our analysis, showing the dynamic asymmetry between a plus and minus defect, which is consistent with the experimental observations of cell density in the vicinity of defects in murine neural progenitor epithelia~\cite{kawaguchi2017topological}. 
	In the bottom panel, we show that this asymmetry in the shape in the neighborhood of $\pm 1/2$ defects is reflected in the Gauss curvature of the surface which is positive (negative) near a plus (minus) defect, consistent with independent observations in a different experiment~\cite{maroudas2020topological}. 
	This correlation between the charge of the defects and the sign of the curvature is a consequence of the relaxational dynamics of the metric even in an active nematic, and as we will see later remains valid even at later (late) times. See \emph{Role of activity for flat configuration} of SI for the effect of activity.

	\begin{figure*}[t]
		\centering
		\subcaptionbox{}
		{	
		\includegraphics[width=.25\linewidth,height=.17\textwidth,trim = {0cm -1cm 0cm 1cm},clip]{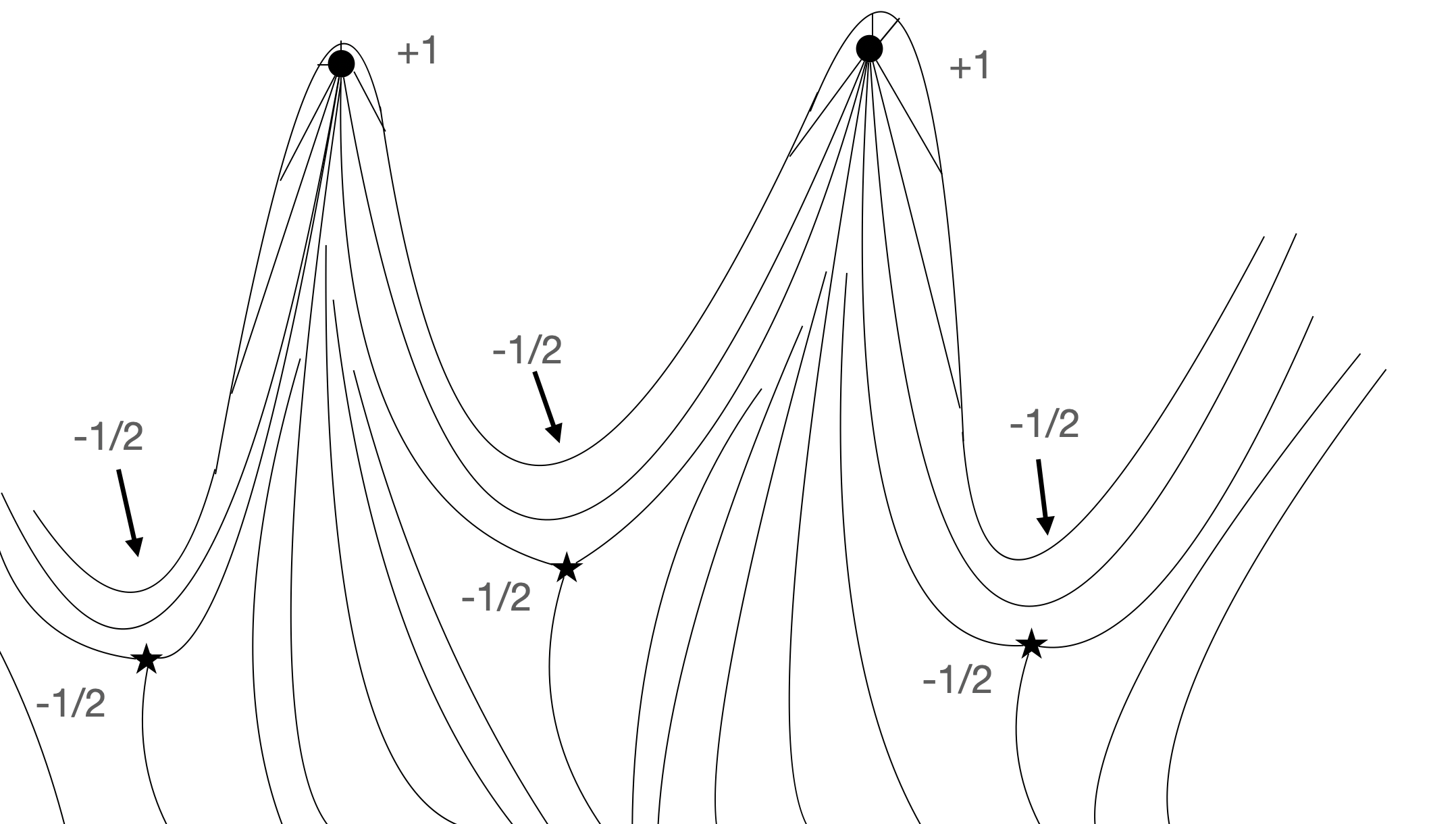}}
		\subcaptionbox{$t=0.004\tau_\varphi$}
		{\includegraphics[width=.15\linewidth]{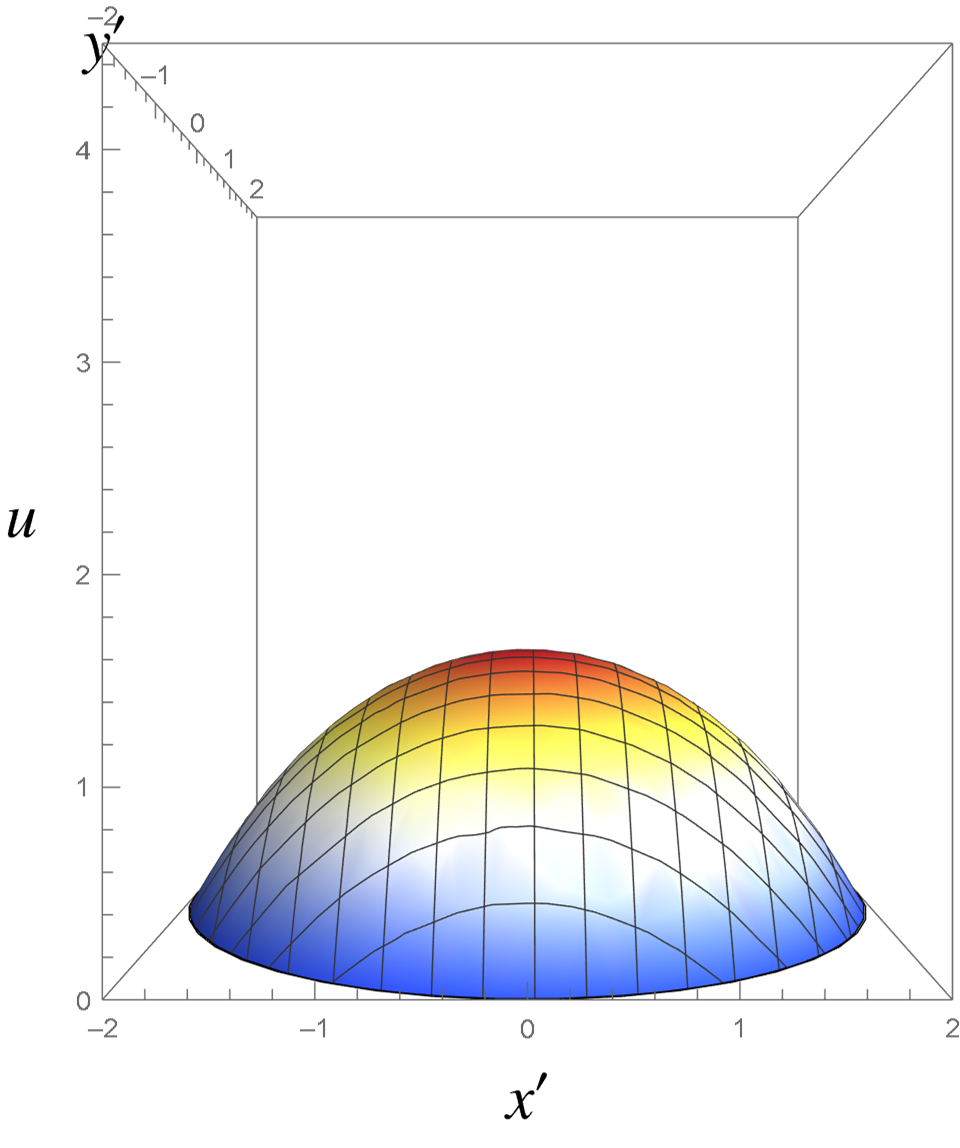}}
		\begin{picture}(0,0)
	    	\put(-40,55){\includegraphics[width=.07\textwidth, height=.07\textwidth]{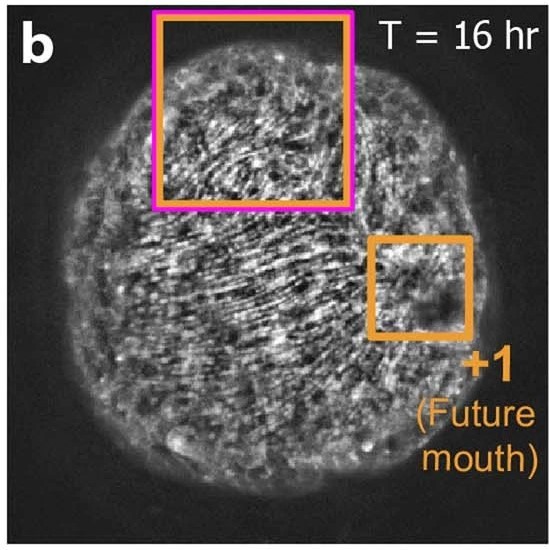}}
		\end{picture}
		\subcaptionbox{$t=0.13\tau_\varphi$}
		{\includegraphics[width=.15\linewidth]{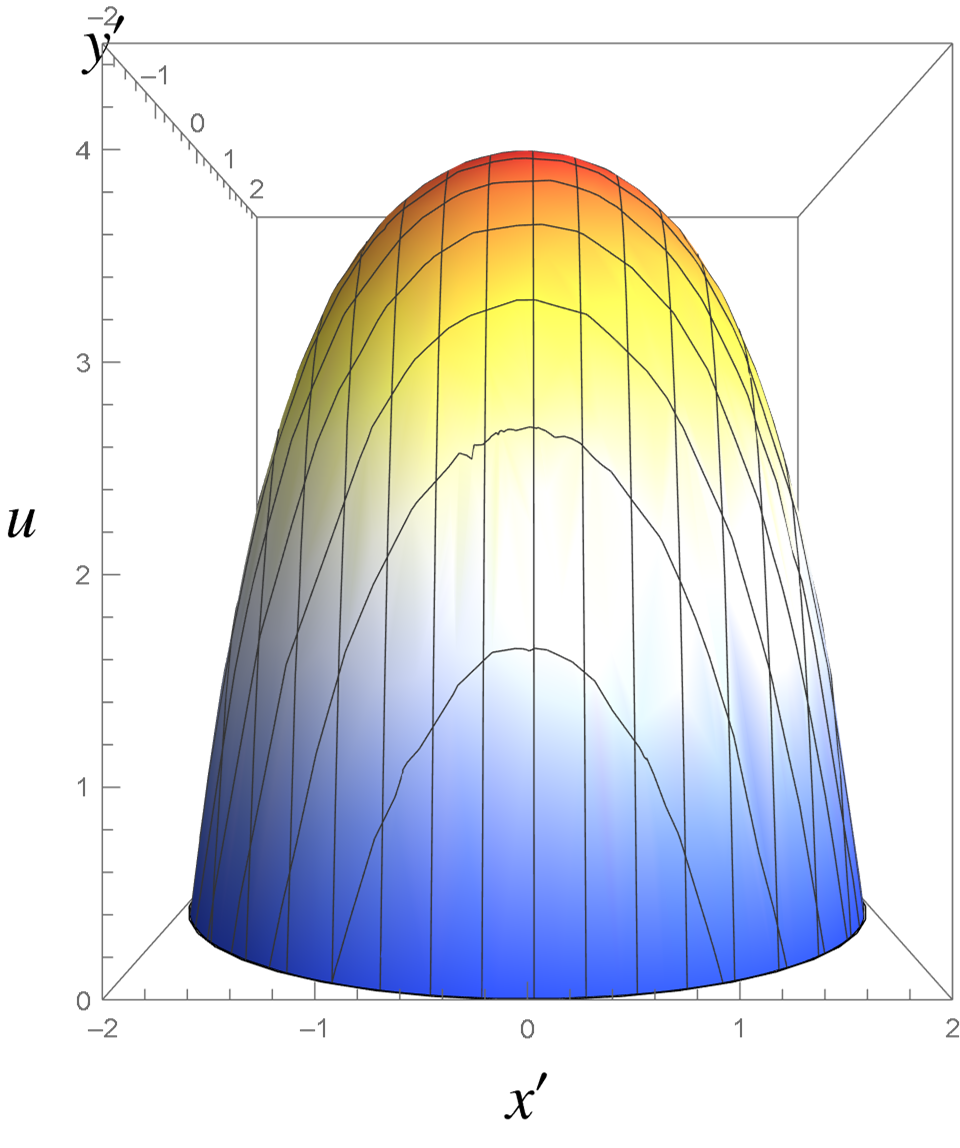}}
		\begin{picture}(0,0)
			\put(-40,55){\includegraphics[width=.07\textwidth, height=.07\textwidth]{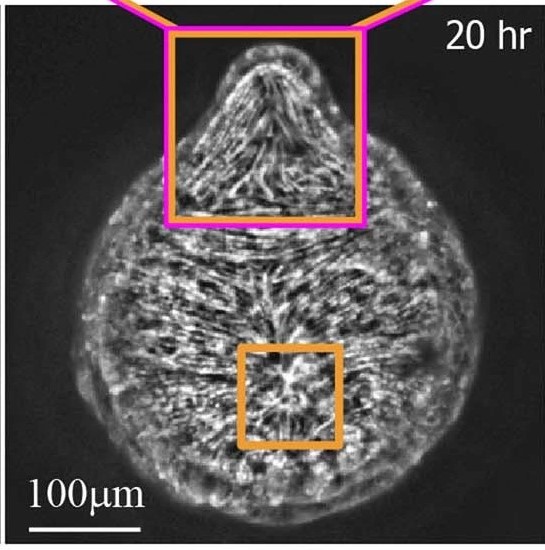}}
		\end{picture}
		\subcaptionbox{}
		{\includegraphics[width=.25\linewidth]{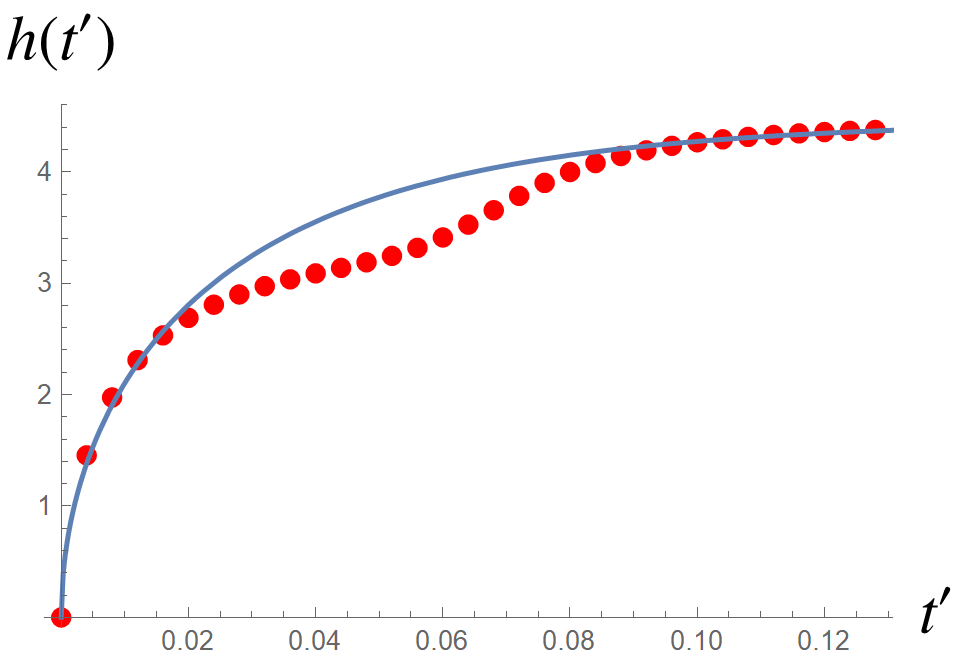}}
		\caption{In (a), sketch of the geometry for the tentacle configuration from our simulation. The black dots represent $+1$ defects, the stars represent $-1/2$ defects, and black lines depict the nematic order. Three of the $-1/2$ defects are on the opposite side. In (b) and (c): snapshots from simulations of height $u$ of tentacle in real space near a $+1$ defect for early and late times, where insets (adapted from~\cite{maroudas2020topological} are snapshots of tentacle formation near a $+1$ defect for early and late times.
			In (d), plot of the height $h(t)$ at the center of the $+1$ defect as a function of time $t$. Red points are data from simulation and blue curve is the fit $h(t) = h_0 [1 - \exp(-t/\tau)]^{1/2}$, where we find that $h_0 = 4.46 L$ and $\tau = 0.04 \tau_\varphi$. Initially, $h(t)\propto \frac{L}{\xi} \frac{L}{\ell_\varphi} L\sqrt{\frac{t}{\tau_\varphi}}$ and $\tau\propto\tau_\varphi$. See \emph{Algorithm for finding embedding} of SI for details.
			All plots use rescaled coordinates $x' = x/L$, $y' = y/L$, and $t' = t/\tau_\varphi$.
		}
		\label{fig:u}
	\end{figure*}
	
	We now turn to consider the effect of activity, i.e. $\alpha  \ne 0$, and rewrite the coupled equations ~\eqref{eq:dQComplexdt}-\eqref{eq:dphidt} in complex coordinates with $\partial_t Q \to D_t Q = \partial_tQ + v^z \nabla_z Q + v^{\bar z} \nabla_{\bar z} Q - (\nabla_z v^z - \nabla_{\bar z} v^{\bar z})Q$ and $\partial_t \varphi \to D_t\varphi = \partial_t\varphi + 2(\nabla_z v^z + \nabla_{\bar z} v^{\bar z})$, where in the over-damped limit, $v^z = \alpha \nabla_z Q = \alpha [\partial Q + 2(\partial \varphi)Q]$ and  $\nabla_z v^z = \partial_z v^z + (\partial\varphi)v^z$. The main effect of the active contribution $-2\vec\nabla\cdot \vec v$ to the initial dynamics of $\varphi$ is that unlike the passive contribution, it is anisotropic, with vanishing average (details in \emph{Role of activity for flat configuration} of SI).
	
	To solve these equations and follow the nematic field and the intrinsic geometry, we use a finite-difference scheme with periodic boundary conditions to simulate a ring-like structure seen in {\it Hydra} (see \emph{Numerical methods} of SI for details). We find that an initial state, defined by a flat geometry and defect configuration of a single $+1$ defect in the center and two $-1/2$ defects on the edges, using the ansatz from~\cite{vafa2020multi-defect}, settles into a stationary defect configuration of a ring of equally spaced $+1$ defects (bound state of two $+1/2$ defects) separated by pairs of $-1/2$ defects in a cylindrical geometry (see Fig.~\ref{fig:latticePlots}(a)), similar to that observed in~\cite{maroudas2020topological}. Activity plays a key role in stabilizing this configuration, and in particular, the $+1$ bound state is a result of balance of Coulombic repulsion force between the defects and motility (see 
	\emph{Topological Defects} of the SI for details). Indeed, the larger the activity parameter for the extensile case $\alpha<0$, the tighter is the $+1$ bound defect. Moreover, the curvature is positive near a plus defect, and negative near a minus defect, as can be seen in Fig.~\ref{fig:latticePlots}(b). 
	Plotting the profiles of $|Q|$ and $\varphi$ along the vertical $y$-axis, we find that the peak in $\varphi$ near the origin indicates outward bulging of the geometry.
	Moreover, the profile of $|Q|$ which is dictated by the nematic coherence length is much narrower than the width of $\varphi$ along the $y$-axis, which is expected given that the geometric coherence length is larger than the nematic coherence length, i.e. $\ell_\varphi > \xi$ and similar to what was observed experimentally in~\cite{kawaguchi2017topological} and in numerical simulations of phase field models e.g. \cite{metselaar2019topology}. 
	
	To ground these results, we turn to observations of epithelial morphogenesis in \emph{Hydra}, a small, fresh-water basal marine invertebrate that has been a model organism for studying the dynamics of body shaping ~\cite{galliot2012hydra,braun2018hydra,maroudas2020topological}. 
	The tubular body of the organism consists of a bilayer of epithelial cells which contains parallel, condensed supracellular actin fibers, which align parallel to the body axis in the outer (ectoderm) layer and perpendicular to the body axis in the inner (endoderm) layer~\cite{aufschnaiter2017apical}. 
	A variable number of tentacles form a ring around the body, near the head, and form when a single $+1$ defect is surrounded by a pair of $-1/2$ defects~\cite{maroudas2020topological}, with the sign of the curvature is correlated with the sign of the defect, consistent with our results summarized in Fig.~\ref{fig:latticePlots}. Indeed, a qualitative rendering of the shape associated with the presence of these bound defect states shown in Fig.~\ref{fig:u}(a) provides a simple projective view of the body plan in the neighborhood of the ring of tentacles.
	
	But what is the actual shape of the epithelium in three dimensions? We note that knowing the intrinsic geometry does not always allow us to deduce the extrinsic geometry, but it is possible to get an approximation by numerically solving the two-dimensional Monge-Amp\`ere equation (details in \emph{Algorithm for finding embedding} of SI). This allows us to estimate the local shape of the active surface shown in Fig.~\ref{fig:u}(b),(c) near a $+1$ defect. Furthermore, we see that at early times the time evolution of the height follows the scaling law $h\propto \sqrt{t}$, which can be analytically derived by using Eq.~\eqref{eq:dphi+dt} (see \emph{Algorithm for finding embedding} of SI for details).
	
	Our minimal framework coupling the dynamics of an active nematic field on a curved surface to the intrinsic geometry of the surface via relaxational dynamics has focused on the interplay between geometry and nematic defects and leads to three simple conclusions: (i) the sign of the curvature is correlated with the sign of the defect, (ii) cells accumulate and form mounds at positive defects and are depleted at negative defects, and (iii) a stationary ring configuration of equally spaced $+1$ defects separated by pairs of $-1/2$ defects can form. These results are consistent with experimental observations in different systems such as neural progenitor cells in-vitro and \emph{Hydra} morphogenesis in-vivo. A more complete description must include a complete characterization of the dynamics of embedding and the possible time-dependence of isothermal coordinates,  as well as feedback on activity of the form $\alpha = \alpha(Q^{\mu\nu}, g_{\mu\nu},\ldots)$, potential directions for future work. We note that a recent preprint \cite{hoffmann2021defect} submitted after the first version of the current paper was submitted has begun to address some closely-related questions.
	
	\acknowledgments
	
	We thank Xianfeng David Gu and Shing-Tung Yau for valuable discussions on reconstructing the embedding from the intrinsic metric. We have also benefited from discussions with Suraj Shankar, and members of M.~Cristina Marchetti's and Mark Bowick's groups. The work of LM is partially supported by the NSF Simons Center for Mathematical and Statistical Analysis of Biology Award No. 1764269.
	
	\bibliography{refs}

	\onecolumngrid
	\clearpage
	\begin{center}
		\textbf{\large Supplementary Material}
	\end{center}
	\appendix
	
	\setcounter{figure}{0}
	\setcounter{section}{0}
	\setcounter{equation}{0}
	\renewcommand{\thefigure}{S\arabic{figure}}
	\renewcommand{\theequation}{S.\arabic{equation}}
	
	 \section{Elementary topological defects and their bound states}
	
	\begin{figure}[t]
		\centering		\subcaptionbox{$+1/2$ defect}
		{\includegraphics[width=0.45\columnwidth]{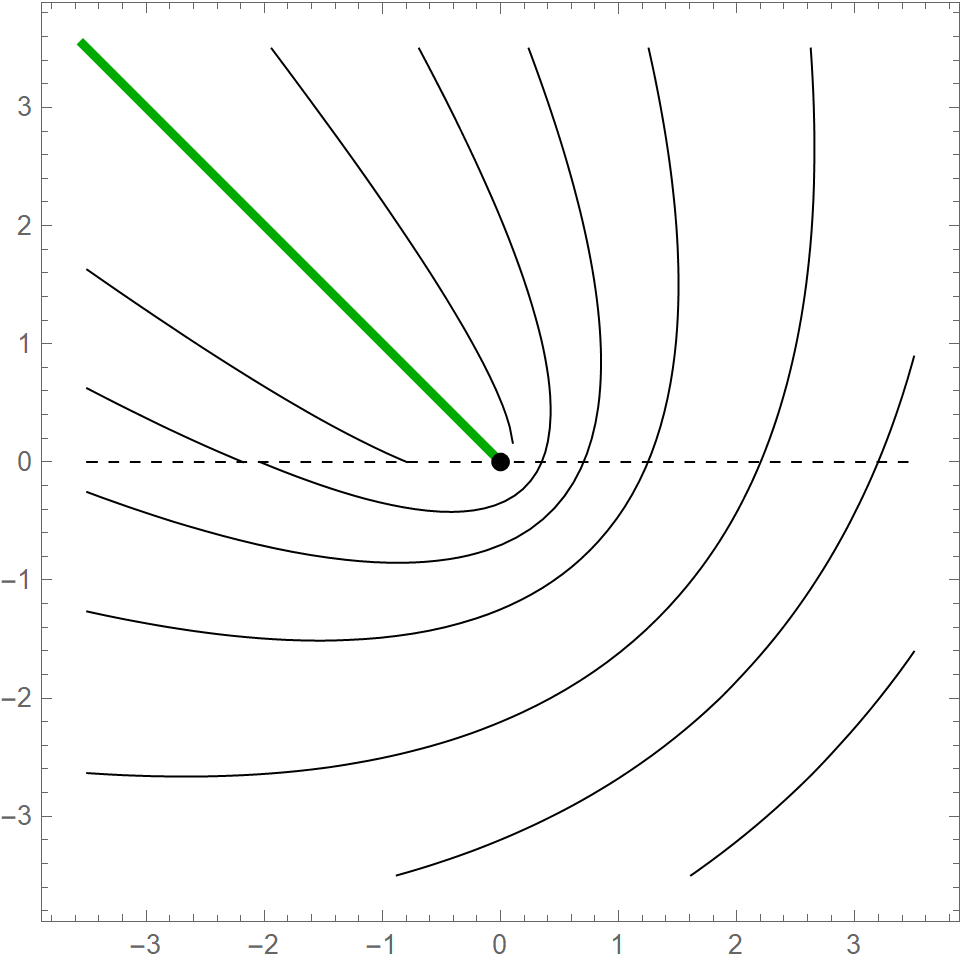}}
		\subcaptionbox{$-1/2$ defect}
		{\includegraphics[width=0.45\columnwidth]{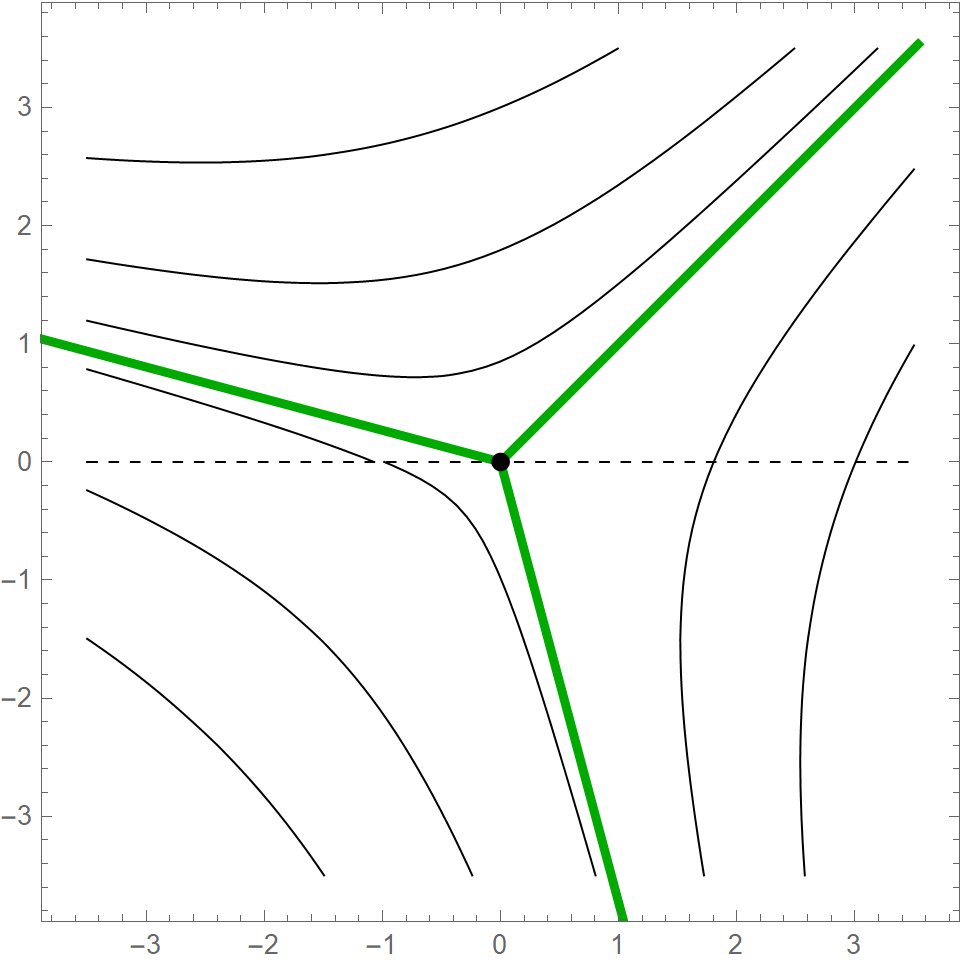}}\\
		\subcaptionbox{$+1$ aster bound state}
		{\includegraphics[width=0.45\columnwidth]{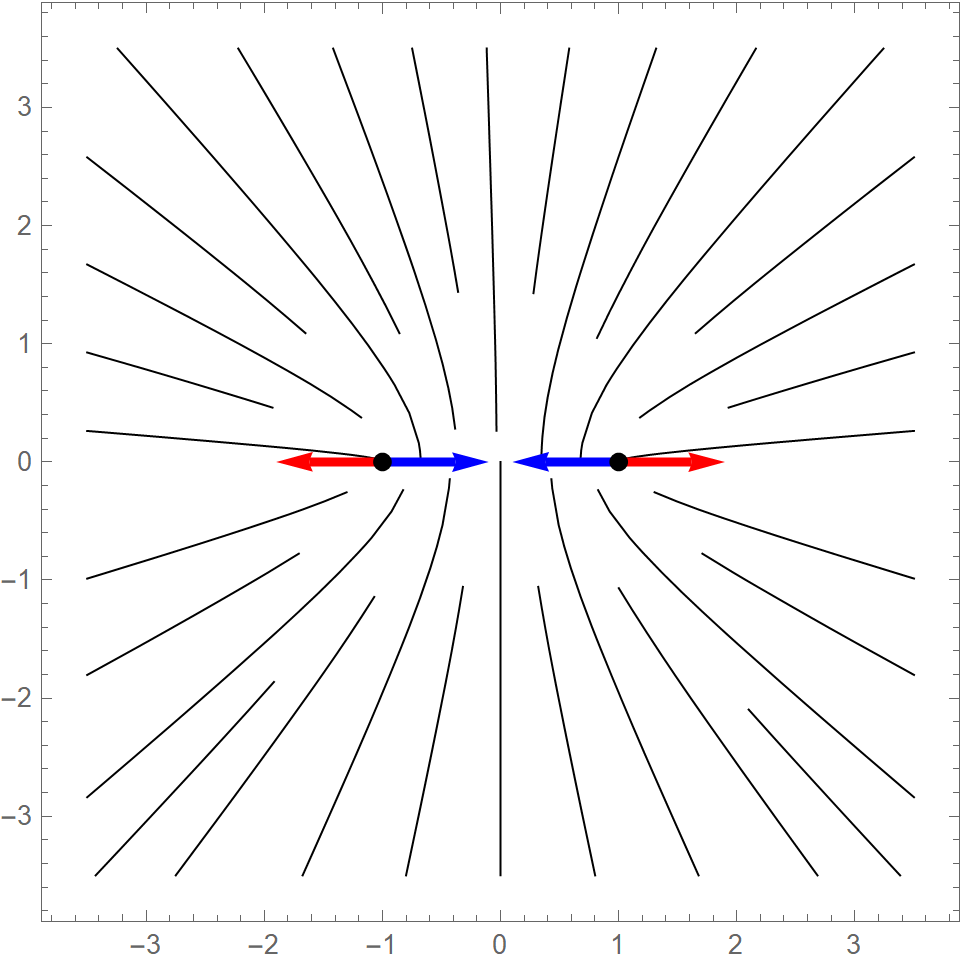}}
		\subcaptionbox{$+1$ vortex bound state}
		{\includegraphics[width=0.45\columnwidth]{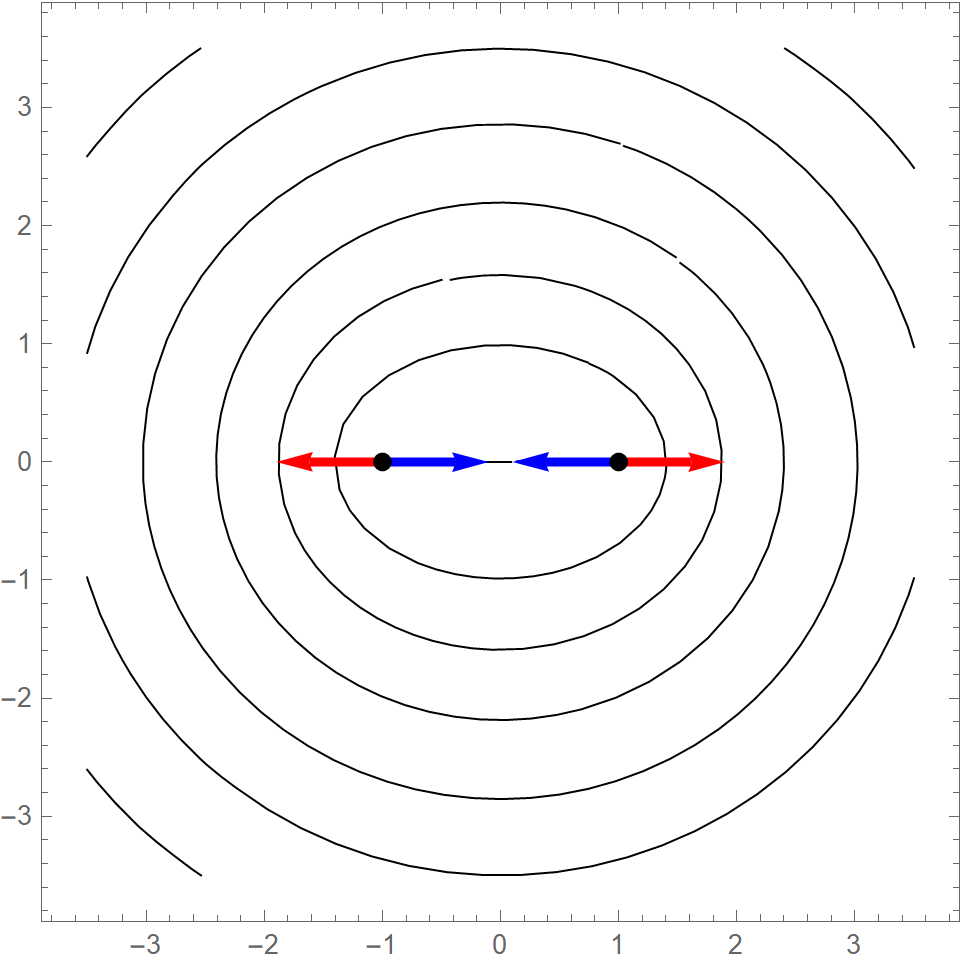}}
		\caption{Top row: sketches of single defect textures for (a) a $+1/2$ defect, and (b) a $-1/2$ defect, where green lines indicate the axes of symmetry. Bottom row: sketches of bound state of two $+1/2$ defects for (c) bound aster state in extensile system and (d) bound vortex state in contractile system, where blue arrows depict motile forces and red arrows depict Coulomb forces.}
		\label{fig:defectStates}
	\end{figure}
	
	The elementary {topological} defects for a nematic are $\pm 1/2$ defects (see Fig.~\ref{fig:defectStates}). An important distinction between the two is that activity renders a $+1/2$ defect motile and capable of self-propulsion along its axis, whereas due to its three-fold symmetry, a $-1/2$ defect remains stationary ~\cite{narayan2007long,sanchez2012spontaneous,giomi2013defect,pismen2013dynamics}. Extensile (contractile) activity drives the $+1/2$ defect to move towards (away from) the head of the comet. Activity can induce a bound $+1$ defect state of two $+1/2$ defects, when the Coulomb repulsion balances the self-propulsion of the $+1/2$ defects; depending on the sign of the activity or equivalently, the orientation of the $+1/2$ defects~\cite{shankar2018defect,thijssen2020binding,vafa2020defect}, one then gets either aster or vortex bound states (see Fig.~\ref{fig:defectStates}).\footnote{There are other forces, such as collective defect drag and activity-induced pair-wise interactions, but we neglect these forces since they are subleading~\cite{vafa2020multi-defect}.}
	
	\section{Role of activity for flat configuration}
	
	Here we will show the the effect of activity on the initial growth, starting from a flat configuration $(\varphi=0)$. We have
	\beq \partial_t \varphi \to \partial_t \varphi + 2\vec\nabla\cdot \vec v\eeq
	so the extra contribution to the RHS of $\partial_t \varphi$ is -2$\vec\nabla\cdot \vec v$. In terms of $Q$,
	\beq 2\vec\nabla\cdot \vec v = 2(\nabla_z v^z + \nabla_{\bar z} v^{\bar z}) = 2\alpha\partial^2 Q + c.c.\eeq
	Substituting $Q^\pm = A(r)e^{\pm i\phi}$ into Eq. S(2) where $\phi$ is the polar angle leads to
	\beq -2\vec\nabla\cdot \vec v = 
	\alpha\begin{cases}
		2\epsilon^{-2}(1 - A^2)A\cos\phi & \sigma = +1/2\\
		-\left(A'' - 3\frac{A'}{r} + 3\frac{A}{r^2}\right)\cos(3\phi) & \sigma = -1/2\\
	\end{cases}
	\eeq
	
	In other words, for the active contribution, unlike the passive contribution, the growth is anisotropic and grows maximally along symmetry axes and least in the opposite directions. Also we want to note that for $\phi=0$, the active contribution for a $+1/2$ defect is always the same sign as the activity, whereas for a $-1/2$ defect, it's the opposite sign. Also in particular, taking the difference between the contributions for $+1/2$ and $-1/2$ defects leads to
	\beq -2(\vec\nabla\cdot \vec v^+ -\vec\nabla\cdot \vec v^-) = -4\alpha\frac{d}{dr}\left(\frac{A}{r}\right)\eeq
	which for $\alpha>0$ is negative for $r \ll 1$ and positive in the limit $r=\infty$.
	
	\section{Numerical methods}
	\label{app:numerics}
	
\begin{figure}[t]
		\centering		
		\subcaptionbox{}
		{\includegraphics[width=0.49\columnwidth]{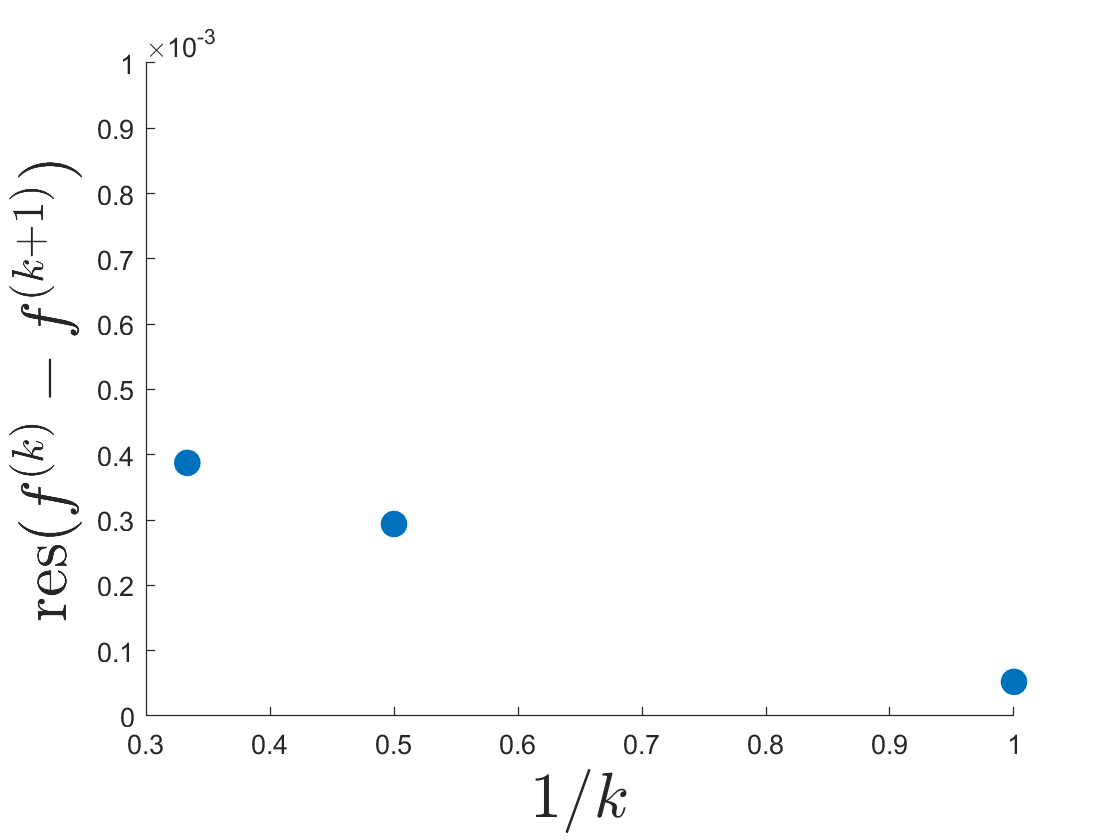}}
		\subcaptionbox{}
		{\includegraphics[width=0.49\columnwidth]{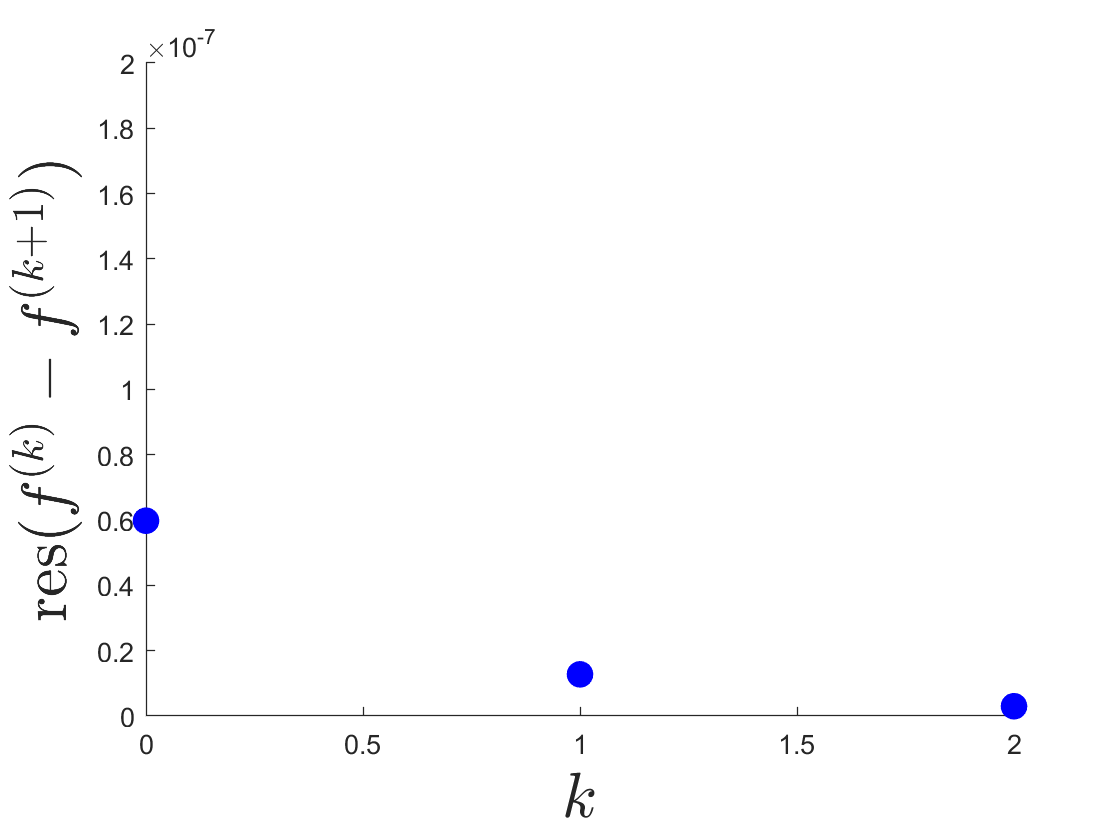}}
		\caption{Convergence checks for (a) discretization $dx$ (lattice spacing) and (b) time step $dt$, in units of $\Delta x$ and $\Delta t$, respectively. In (a), the number of points in each spatial direction is scaled by $1/k$, and in (b), the total number of time-point scales with $k$ as $N 2^k$. In both plots, $f^{(k)}$ denotes $(|Q^{(k)}|,\varphi^{(k)})$ for the simulation with the total number of points corresponding to $k$. As expected, the residual error decreases for higher spatial and temporal resolution.}
		\label{fig:checks}
	\end{figure}

	We simulated equations~(9)-(10) (with the substitution $\partial_t \to D_t$) with a custom code written in Matlab. 
	Our domain of size $40\times 120$ with periodic boundary conditions is represented by a grid of size $100\times300$ with spatial discretization $\Delta x = \Delta y = 1$. Spatial derivatives are evaluated using second-order central difference scheme. The initial condition is flat geometry ($\varphi=0$) and defect configuration of a single $+1$ defect in the center (which due to periodic boundary condition forces the appearance of two $-1/2$ defects), using the ansatz from~\cite{vafa2020multi-defect}. 
	The equations are time-integrated using a forward Euler finite-difference method with time-step $\Delta t = 0.002$, for a total number of 80,000 iterations (until convergence to stationary solution). Convergence checks for the spatial and temporal resolution are in Fig.~\ref{fig:checks}.
	
	We then extracted the final data (restricted to a region of size $10\times 10$ where the curvature is strictly positive) and after smoothing it via Mathematica's GaussianFilter function with standard deviation $\sigma = 15$, we used Mathematica's NDSolve function to numerically solve and thus find the embedding (plots in Fig.~5 of the main text) according to the iterative algorithm described in the next section.

	\section{Algorithm for finding embedding}
	\label{app:embedding}
	
	To determine the extrinsic geometry from the intrinsic geometry, we can make some progress by noting that if  $X^i = (x,y,u(x,y))$ is the Monge representation of the surface in 3D, the Gauss curvature $R$ is given by
	\beq R(x,y) = \frac{u_{xx} \cdot u_{yy} - u_{xy}^2}{(1 + u_x^2 + u_y^2)^2} \; .\eeq
	In terms of $R$, noting that $(\Delta u)^2 = u_{xx}^2 + u_{yy}^2 + 2u_{xy}^2 + 2R(1 + |\nabla u|^2)^2$ allows us to determine $u$ for positive $R$ by an iterative procedure\footnote{We would like to thank Xianfeng David Gu for explaining this method.}, $u^{(n+1)} = \mathcal P[u^{(n)}] \; $ (see~\cite{benamou2010two,gu2013variational,cui2019spherical} and references therein, and \cite{guGeometric} and \cite{guSpherical} are accompanying open source codes for the latter two references), where
	\beq\mathcal P[u] = \Delta^{-1} \sqrt{u_{xx}^2 + u_{yy}^2 + 2u_{xy}^2 + 2R(1 + |\nabla u|^2)^2} \;. \eeq Expressing $u$ as
	$u = u_0 + \delta u$, where $u_0= \frac{\beta}{2}(x^2 + y^2)$ is our initial seed for the iterative procedure, and $\delta u$ is a small correction, with $\beta$ chosen so that $\beta^2 \sim R$ at the peak, we can determine $u$ for representative times, as shown in Fig.~5. For our data, we did three such iterations.
	
	\section{Dynamics of bud growth}
	\label{app:bud}
	We now present a heuristic argument for the dynamics of budding.
	From Eq.~(11), at the center of the $+1$ defect, \beq\partial_t\varphi^+ \sim \gamma_\varphi^{-1} K |\partial Q^+|^2 \sim \gamma_\varphi^{-1}\epsilon^{-2} = (L/\ell_\varphi)^2\tau_\varphi^{-1}\; ,\eeq
	and so by taking the Laplacian, 
	\beq \partial_t R \sim  \gamma_\varphi^{-1} K \nabla^2 |\partial Q^+|^2 \sim (L/\ell_\varphi)^2\tau_\varphi^{-1}\xi^{-2} \eeq
	so that 
	\beq R\sim (L/\ell_\varphi)^2(t/\tau_\varphi) \xi^{-2} \; . \eeq
	Combining this with the fact that at the center of a $+1$ defect core, $h\propto\beta L^2$, where $L$ is the separation between neighboring positive and negative defects, leads to 
	\beq h^2 \propto RL^4 \propto \left(\frac{L}{\xi}\right)^2\left(\frac{L}{\ell_\varphi}\right)^2L^2\frac{t}{\tau_{\varphi}} \; . \eeq
	
	Since the nematic order is a fast variable, we can assume it settles before the (slower) geometric relaxation. The leftover dynamics is due to the diffusive nature of the Ricci flow (essentially the heat equation), leading to $(\varphi - \varphi_\text{eq}) \to \exp[-t/\tau]$, where $\tau\sim\tau_\varphi$. Similarly, this leads to $(h - h_0) \to \exp[-t/\tau]$.
	
	These asymptotics are consistent with the fit we found from our simulations in Fig.~5,
	\beq h(t)^2 = h_0^2\left(1 - e^{-t/\tau}\right)\label{eq:hSq} \; ,\eeq
	if we take $h_0\sim \left(\frac{L}{\xi}\right)\left(\frac{L}{\ell_\varphi}\right)L$ and $\tau\sim\tau_\varphi$.
	
	We thus see that the relative size of the  tentacle $h_0$ to the average tentacle separation $L$ is 
	\beq\frac{h_0}{L} \sim \left(\frac{L}{\xi}\right)\left(\frac{L}{\ell_\varphi}\right) \; . \eeq so that a tentacle can grow in two different ways for fixed tentacle separation ($L$): by decreasing either the nematic ($\xi$) or geometric ($\ell_\varphi$) coherence length.
	
\end{document}